\documentclass[%
reprint,
superscriptaddress,
amsmath,amssymb,
aps,
prb,
]{revtex4-1}

\usepackage{graphicx}
\usepackage{bm}
\usepackage{enumerate}

\begin{document}

\title{Thermodynamics of the classical spin-ice model with nearest neighbour interactions using the Wang-Landau algorithm}

\author{M. V. Ferreyra}
\affiliation{Instituto de F\'{\i}sica de L\'{\i}quidos y Sistemas Biol\'ogicos  (IFLySIB), UNLP-CONICET, 1900 La Plata, Argentina}
\affiliation{Facultad de Ciencias Exactas y Naturales, Universidad Nacional de La Pampa, 6300 Santa Rosa, Argentina.}

\author{G. Giordano}
\affiliation{ Departamento de F\'{\i}sica, Facultad de Ciencias Exactas, Universidad Nacional de La Plata, 1900 La Plata, Argentina}

\author{R. A. Borzi}
\affiliation{Instituto de F\'{\i}sica de L\'{\i}quidos y Sistemas Biol\'ogicos  (IFLySIB), UNLP-CONICET, 1900 La Plata, Argentina}
\affiliation{ Departamento de F\'{\i}sica, Facultad de Ciencias Exactas, Universidad Nacional de La Plata, 1900 La Plata, Argentina}

\author{J. J. Betouras}
\affiliation{Department of Physics, Loughborough University, Loughborough LE11 3TU, UK}

\author{S. A. Grigera}
\affiliation{Instituto de F\'{\i}sica de L\'{\i}quidos y Sistemas Biol\'ogicos  (IFLySIB), UNLP-CONICET, 1900 La Plata, Argentina}
\affiliation{School of Physics and Astronomy, University of St Andrews, St Andrews, KY16 9SS, United Kingdom}

\date{\today}

\begin{abstract}
In this article we study the classical nearest-neighbour spin-ice model (nnSI) by means of Monte Carlo simulations, using the Wang-Landau algorithm.  The nnSI describes several of the salient features of the spin-ice materials. Despite its simplicity it exhibits a remarkably rich behaviour.  The model has been studied using a variety of techniques, thus it serves as an ideal benchmark to test the capabilities of the Wang Landau algorithm in magnetically frustrated systems.  We study in detail the residual entropy of the nnSI and, by introducing an applied magnetic field in two different crystallographic directions ([111] and [100],) we explore the physics of the kagome-ice phase, the transition to full polarisation, and the three dimensional Kasteleyn transition. In the latter case, we discuss how additional constraints can be added to the Hamiltonian, by taking into account a selective choice of states in the partition function and, then, show how this choice leads to the realization of the ideal Kasteleyn transition in the system.
\end{abstract}

.
\maketitle

\section{Introduction}

In magnetism, when a spin cannot fully minimise its interactions with its neighbours, the system is called ``frustrated''.  This situation can arise under a number of different circumstances, such as bond disorder, further neighbour interactions and lattice geometry. Geometrically frustrated magnets are the cleanest and better controlled in experimental systems.  Frustration precludes the formation of simple ordered ground states, rather, it typically leads to a degenerate manifold of ground states, scaling with the system size, therefore, exhibiting extensive zero-point entropy.  Unsurprisingly, these ground states are unstable to any small perturbation. Frustrated systems exhibit a rich variety of behaviour, including order by disorder, fractionalisation and magnetic analogues of solids, liquids, glasses, ice, quantum liquids and bose condensation.  They represent ideal model systems for the study of generic concepts relevant to collective phenomena, where simple classical Hamiltonians can give rise to a wealth of different phenomena \cite{Ramirez1994,Moessner2001,Lacroix2011}.  Given that analytical treatment cannot be performed, frustrated magnets are very well suited for numerical modelling, and constitute ideal test-grounds for powerful numerical techniques.

Among the recent developments in classical simulation techniques, the Wang-Landau algorithm (WLA) has stood out  as a powerful tool for the determination of phase diagrams \cite{WangLandau2001-I,WangLandau2001-II}.  The aim of this  technique is to give a very accurate estimate of the density of states of the system (DoS), from which the thermodynamic behaviour can be calculated using canonical ensemble statistical mechanics.   There are two particular points of this technique that makes it attractive for the study of frustrated systems: first,  by simply providing an estimate of the DoS it gives direct information about the degeneracy of the ground state and  its low temperature excitations; second,  in its algorithmic construction it has a built in mechanism to avoid the trapping of the system into local minima in the energy landscape.  The existence of many local minima in frustrated magnets is one of the main reasons of the under-performance of the usual Monte Carlo techniques.  Recently, the WLA has been successfully applied to specific problems in frustrated systems and dimer models, such as the formation of magnetisation plateaus in the Ising Shastry-Sutherland model \cite{Wslin2014} or to determine the order of a transition in the Heisenberg stacked triangular antiferromagnet \cite{Ngodiep2008} and in dimer models with next-nearest-neighbor interactions \cite{papanikolaou2010}. In the present work we explore thoroughly a simple classical frustrated model, the nearest-neighbour spin-ice model (nnSP), by means of the WLA. This model, despite its apparent simplicity ---Ising spins on a pyrochlore lattice interacting ferromagnetically at nearest neighbours--- exhibits an extremely rich behaviour.  The degeracy of the zero field ground state grows exponentially with the system size and with aufbau rules analogous to those of protons in water ice.  By applying an external magnetic field it is possible to find a metamagnetic transition or to tune into an effective two dimensional model (the kagome-ice, also with extensive residual entropy) and to find two- and three-dimensional Kasteleyn transitions. The physics of this model system has been well studied using a variety of techniques  \cite{denHertog2000,Isakov2004,Moessner2003,Jaubert2008,Jaubert2009}, and thus serves as an ideal benchmark to test the capabilities of the WL algorithm in this type of systems.  Beyond that, the WLA provides new thermodynamic results that have not been obtained by other methods, such as the free energy as a function of the order parameter, which, to our knowledge, has been calculated for a Kasteleyn transition for a first time.

The structure of this article is as follows. In the next section we give a brief discussion of the model and the simulation technique for completeness. In section III, we study the model at zero applied field, calculate the residual entropy and compare it with different existing estimates using the WLA technique. In the same section, we study the system under field applied along [111], and explore the entropy of the kagome-ice state, as well as the entropy peak that rises when this state is destroyed by increasing the temperature.  Finally, we study the case of field applied along [100], and do a characterisation of the three dimensional Kasteleyn transition into the fully polarised state.

\section{Methods}
\subsection{Wang Landau Algorithm} 

Recently, F. Wang and D. P. Landau introduced an algorithm to estimate the density of states of a system by performing a random walk in energy space \cite{WangLandau2001-I,WangLandau2001-II}.  This method is closely related to `umbrella sampling' techniques \cite{Torrie1977} and multicanonical Monte Carlo \cite{Berg1992}. The algorithm provides a very good estimate of the DoS of the system over a bounded region of the energy spectrum.  The DoS can be calculated as a function of the energy, if one works in the canonical ensemble, but also as a function of other variables like pressure or magnetisation if one is interested in other ensembles such as the isothermal-isobaric or its magnetic equivalent (as we will use in sections \ref{h111} and \ref{h100}).  In this section we describe the procedure for one variable, which is then straightforwardly extended to the case of several variables.

The algorithm requires the knowledge of the Hamiltonian of the system and a method to sample configurations - in our case it is simply random single spin flips. The starting point is an arbitrary initial configuration with energy $E_0$.  Initially, the DoS is taken as homogeneous: $g(E)=1$ for all $E$.  One step of the calculation consists then in choosing new random configuration, calculating its energy $E_1$, and accepting it or discarding it with probability
\begin{equation}
	p(E_0\rightarrow E_1)= \min (1,\frac{g(E_0)}{g(E_1)}).
    \label{prob}
\end{equation}
\noindent
At each step, the histogram $H(E)$ and the DoS $g(E)$ of the final configuration are modified according to $g(E) \rightarrow g(E)f$ and $H(E) \rightarrow H(E)+1$. Initially the modification factor $f$ for $g(E)$ is taken from a larger value (usually, $f_0=e^1$) which is then reduced as the algorithm progresses.  The rule by which the modification factor $f_i$ is reduced is an important choice that conditions the accuracy of the DoS and speed at which the process converges.  In principle, one can choose any function that tends monotonically to one, and stop the process once $f$ reaches a given value.   In the original work, Wang and Landau propose the rule $f_{k+1}=\sqrt{f_k}$, but other choices are possible which give faster convergence.  In particular, Belardinelli and Pereyra (BP) \cite{Bellardinelli2007} proposed and alternative method with improved convergence times, in which the modification factor is eventually scaled as the inverse of the Monte Carlo time, $t$. For this work we have adopted the BP algorithm.

In our case, within one Monte Carlo step this procedure was repeated with single spin flips until the spin of each site had been chosen to change at least once on average. In practice, the quantity $\ln{g(E)} \rightarrow \ln{g(E)} + \ln{f}$ is used for simplicity (hence the choice $f_0=e$).  Notice that this procedure guarantees that the random exploration in phase space will not jam at local minima: each time the transition to the new state is not accepted, the DoS of the initial state is increased and thus the probability of accepting any future transition, which will be proportional to $g(E_0)$, is also increased.  The detailed balance condition is satisfied to within $\ln f$ accuracy.

During this random walk through phase space, the histogram is accumulated and checked periodically. When $H(E)$ becomes \textit{flat}, the histogram is reset, and the next random walk begins with a finer modification factor $f_1$. The criterion of flatness varies according the size and complexity of the system. Usually, the criterion used for flatness is that every entry in $H(E)$ is not smaller than a percentage of the average histogram for all $E$.

The final result is a relative DoS.  To calculate the absolute values one needs some additional information, e.g. a known point in the DoS -- in our case the high temperature value of the DoS tends to $2$, or the knowledge of the integral of the DoS --in our case $2^N$.
Once the DoS of the system is known, it is straightforward to calculate the thermodynamic quantities in the canonical ensemble: for example, from $Z= \sum_i g(E_i) e^{-E_i/k_B T}$, the free energy can be written as $F(T) = -kT \log{Z}$, the internal energy $ U(T)= \frac{1}{Z} \sum_E E g(E)e^{-\beta E}$, the entropy $S(T) = \frac{U(T)-F(T)}{T}$, and the specific heat, using the usual linear fluctuation relation, $C(T) = \frac{\langle E^2 \rangle_T - (\langle E \rangle_T)^2}{T^2}$.

\subsection{Spin ice}

The nnSI Hamiltonian under an external magnetic field reads:

\begin{equation}
H = J_{\rm{eff}} \sum_{<ij>} \textbf{S}_i  \cdot \textbf{S}_j - g \mu_B \sum_{i} \textbf{H} \cdot \textbf{S}_i, \label{HnnSI}
\end{equation}
\noindent where the $\textbf{S}_i$ are Ising spins situated at the corners of a pyrocholore lattice (see inset of Fig.\ \ref{DoS}), $\textbf{H}$ is the external magnetic field, $g$ the gyromagnetic ratio and  $J_{\rm{eff}}$ is the effective exchange interaction which is taken to be positive.

The model is applicable to a certain class of materials, of which the most notable examples are Dy$_2$Ti$_2$O$_7$ and Ho$_2$Ti$_2$O$_7$. In these materials, the magnetic ions sit at the corners of a pyrochlore lattice and are constrained by the crystalline field to point along the local $\langle$111$\rangle$ quantization axes ({\em i.\ e.\ } to or from the centre of the tetrahedra).  The nnSI model provides a very good description of these materials between 0.2K and 10K \cite{Bramwell2004}.  The main difference arises from the fact that in the materials, in addition to the exchange interaction -- which is antiferromagnetic-- there is a large long range dipolar interaction.  If the latter is truncated beyond the nearest-neighbour spins, $J_{\rm{eff}}$ is effectively ferromagnetic \cite{Bramwell2001}.

The ground state of the nnSI scales exponentially with system size and obeys the local construction rule that within any tetrahedra, two of its spins should point inwards, and two outwards \cite{Bramwell2001,Bramwell2004}. This rule is called the ``ice-rule'' given its analogy to the Bernal and Fowler rules for protons in water-ice \cite{Bernal1933} and is the origin of the epithet ``spin-ice'' given to these models. The exponential degeneracy of this ground state leads to an extensive residual entropy at zero temperature.

We analysed the nnSI model using the WLA with both the original implementation and the one proposed by Belardinelli and Pereyra.  In order to reach larger lattices we also performed simulations dividing the energy range in multiple regions. To normalize the DoS, we generally used the condition that there are only two states with maximum energy (the ``all-in'' and ``all-out'' configurations). The comparison between several runs performed with different seeds, or different normalisation, allowed us to estimate the errors on the residual entropy for different sizes. We explored the configuration space through random single spin-flip moves.  In addition, we used a conventional cubic cell for the pyrochlore lattice, which contains 16 spins, and simulated systems with $L \times L$ cells, with $L$ ranging from 1 to 8 (16 to 8192 spins).  The DoS was estimated as a function of energy with the modification factor changing from $f_0=e^1$ to $f_{\rm final}=\exp ({10^{-9}})$.

For the results of the section \ref{heq0} we used the single-variable WLA, while for sections \ref{h111} and \ref{h100} we accumulated the DoS as a function of both the energy and the magnetisation in the direction of the applied magnetic field.  In this latter case, the thermodynamic quantities are calculated using a partition function which is summed in the energy and in a variable conjugate to the magnetic field, in this way the derivatives of $Z$ provide information about the average magnetisation. We have chosen $J=1.11K$ in order to match the effective nn exchange constant for Dy$_2$Ti$_2$O$_7$.

\section{Results}
\subsection{Spin-ice with no applied external field: residual entropy \label{heq0}}

\begin{figure}[!]
\centerline{\includegraphics[width=\columnwidth]{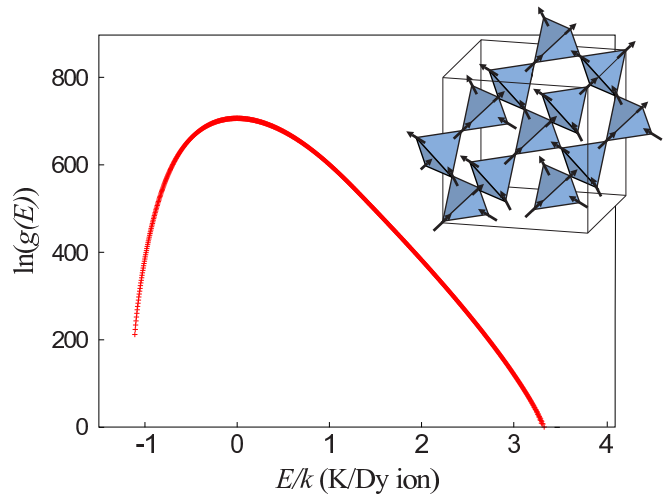}}
\caption{Natural logarithm of the density of states of the nnSI model for a system size $L=4$.  Its shape is characteristically asymmetric with high degeneracy of the ground state. The inset shows an schematic view of the pyrochlore lattice.}
\label{DoS}
\end{figure}

As a first step we analysed the nnSI model with no applied external field ($H=0$), in particular, we were interested in the behaviour of the specific heat and the entropy.  In Fig.\ \ref{DoS} it is shown the logarithm of the DoS of the nnSI model for a system size $L=4$.  Its shape, in contrast with the highly symmetrical DoS of the Ising model (see e.\ g.\ \onlinecite{WangLandau2001-II}), is characteristically asymmetric, starting from a high value at the lowest energy, a direct measure of the high degeneracy of the ground state.

%
%

\begin{figure}[!]
\centerline{\includegraphics[width=\columnwidth]{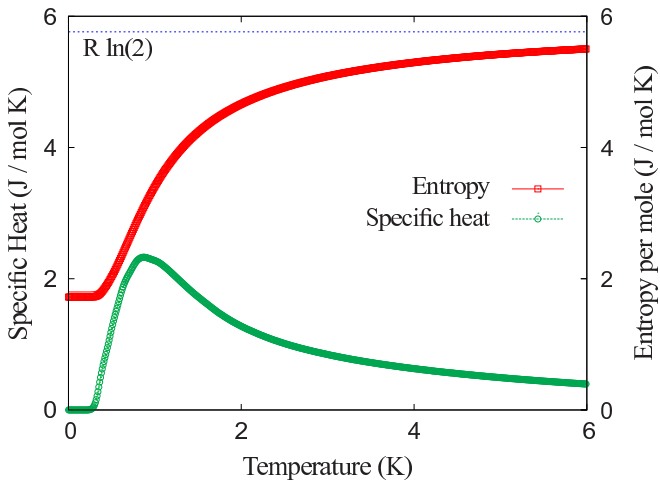}}
\caption{Specific heat (green) and entropy (red) vs temperature for system size L = 4, calculated from the density of states (DoS) with parameters set for Dy$_2$ Ti$_2$ O$_7$. The figure shows the expected Schottky-like peak in the specific heat, marking the onset of spin-ice correlations.  The entropy does not vanish as the temperature is lowered towards $T=0$ but shows the residual value characteristic of all ice models.}
\label{CSvsT}
\end{figure}

In the case of the specific heat, the expectation is that as the system is cooled down, there will be an onset of correlations as the temperature $T$ approaches $J/k$, that will be evidenced by a Schottky-like peak \cite{Bramwell2004}.  The specific heat calculated using the WLA shows a peak close to $T=1K$ (green crosses in figure \ref{CSvsT}), in accordance with those calculated by other techniques \cite{denHertog2000} and with the high temperature experimental results in spin-ice materials ($T>0.6{\rm K}$) \cite{Ramirez1999}.  This same figure shows the temperature dependence of the entropy per mole of the system. The high temperature value (not shown in the figure) is $R\ln 2 \approx 5.76~{\rm J/molK}$, indicating that the system is behaving as an uncorrelated paramagnet.  As the temperature is lowered the entropy is reduced until it reaches a residual value $S_0$ close to $1.7~{\rm J/molK}$.

%
%

The residual entropy is a characteristic feature of ice models; in real spin-ice materials, such as Dy$_2$Ti$_2$O$_7$ or  Ho$_2$Ti$_2$O$_7$ it is expected that the degeneracy of the spin-ice manifold will be lifted by additional interactions ---chiefly the dipolar interaction--- and that the system at $T=0$ will be ordered \cite{Melko2004,Pomaranski2013}.  In the nnSI model, however, the aufbau rules are strictly those of Bernal and Fowler, and one expects to find an exponentially degenerate state with the same extensive residual entropy of the three dimensional ice models.  The determination of the value of this entropy has a long history, starting with Linus Pauling's famous estimate in 1935 \cite{Pauling1935}.  $S_0$ can be written as
\begin{equation}
S_0=k\ln W_N
\end{equation}
\noindent where $k$ is Boltzmann's constant and $W_N=W^{N_T}$ is the number of microstates that form the ground state of the system, with $N_T$ the number of tetrahedra.  Pauling's estimate gives $W=3/2$, which translated to the entropy per mole gives $S_0=R/2\ln 3/2 \approx 1.68~{\rm J/molK}$, in reasonable agreement with the results of Fig \ref{CSvsT}.

\begin{figure}[ht]
\centerline{\includegraphics[width=\columnwidth]{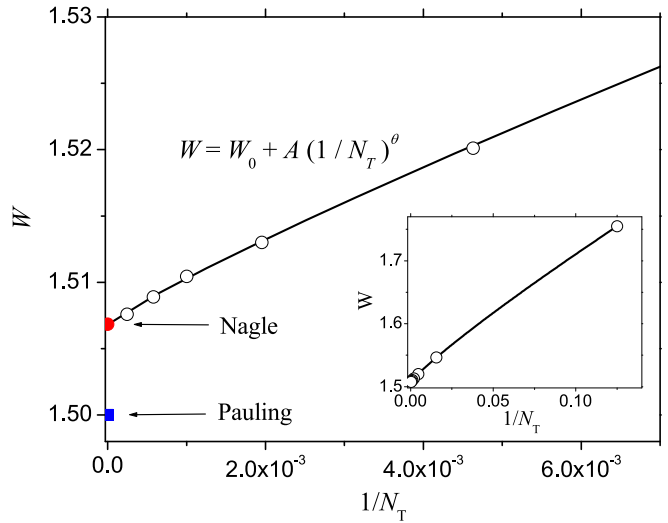}}
\caption{The number of microstates of the ground state,$W$, as a function of the inverse of the number of tetrahedra $1/N_T$. As expected, the residual entropy decreases as the system size is increased. Shown for reference are the thermodynamic value of $W$ from Pauling's estimate (blue square) and from Nagle's calculation (red circle).  The line is a fit according to eq.\ \ref{win}.  The inset shows the full range of the fit (from $L=1$ to $L=8$).}
\label{S0vsN}
\end{figure}

Pauling's estimate neglects correlation effects (in the form of loops) and it can be shown that it is a lower bound on the true $S_0$ \cite{Petrenko1999}. While an exact solution exist for two dimensional ice models \cite{Lieb1967}, this is not true for three dimensions.  Currently the best estimate of the entropy is that due to J.\ F.\ Nagle \cite{Nagle1966}, who, building on work by Stillinger and Di Marzio \cite{DiMarzio1964}, used a series expansion method to derive the estimate
\begin{equation}
 W_{\rm Nagle}=1.50685 (15). \label{Wnagle}
\end{equation}
\noindent The most accurate calculation of $S_0$ for the nnSI model in the literature comes from the integration of energy and magnetisation data obtained by loop Monte Carlo simulations \cite{Isakov2004} and gives $W_{I}=1.5071 (3)$, very close to Nagle's result.  The WLA provides a direct determination of the entropy without the need of integrating the specific heat and specifying additional constants, and is thus ideally suited for an accurate determination of the residual entropy of the nnSI model.  A variant of the WLA has been successfully applied to determine the residual entropy of two simple nearest neighbours ice models in three dimensional hexagonal lattices: the six-state H$_2$O molecule model and the two-state H-bond model \cite{Berg2007}.  A similar method is applied to nnSI.
We have calculated the DoS for lattice sizes $L=1$ to $8$ which correspond to $N_T = 8$ to $4096$ tetrahedra, and from those determined $W$ for the ground state.  Figure \ref{S0vsN} shows $W$ as a function of the inverse of the number of tetrahedra $1/N_T$. We have used two different criteria to normalise the DoS: in one case we used a known density for a given state (the highest energy state) and in the other we used the sum of all states ($2^N$ in our system).  The differences in $W$ given by the different criteria of normalisation, or that obtained in independent runs using a different set of random numbers, are smaller than the size of the symbols in the figure.
%
%

Fig. \ref{S0vsN} shows that, as expected, the residual entropy decreases as the system size is increased.  In order to obtain the thermodynamic value of $W$ we fit this data to the form
\begin{equation}
W(x) = W_\infty + a_1 \left(\frac{1}{N_T}\right)^\theta.
\label{win}
\end{equation}
\noindent From this fit we obtain $W_\infty = 1.50682(9)$, with $a_1=1.557(9)$ and $\theta=0.883(3)$.  The sub-linear value of $\theta$  is an indication of bond correlations in the ground-state manifold.  Our value for $W$ in the thermodynamic limit is perfectly consistent with the results by Nagle (see eq.\ \ref{Wnagle}) and by previous calculations \cite{Isakov2004,Berg2007}.


\subsection{Spin-ice: Field along [111] \label{h111}}

A remarkable feature of the SInn model is that an external magnetic field can tune the system into regions of different physics.  Two notable cases happen when the field is oriented (i) along the crystallographic [111] direction, which will be described in this subsection, and (ii) along [100], which will be described in the following subsection. To describe the effects of an applied external field in the WLA,on one hand the Zeeman term must be included in the Hamiltonian (the second term in eq.\ \ref{HnnSI}) and, on the other hand, it is more convenient to calculate the DoS as a function of two indices: energy and magnetisation $M$, the latter being the quantity conjugate to the field.  In this way is it possible to work in the magnetic equivalent of the isothermal-isobaric ensemble and obtain directly from the doubly summed partition function the Gibbs free energy $G(H,T)$ and the mean value of $M$.  The disadvantage of this approach is that it is too expensive in calculation resources and, therefore, the calculations are usually constrained on smaller system sizes.

%
\begin{figure}[ht]
\centerline{\includegraphics[width=\columnwidth]{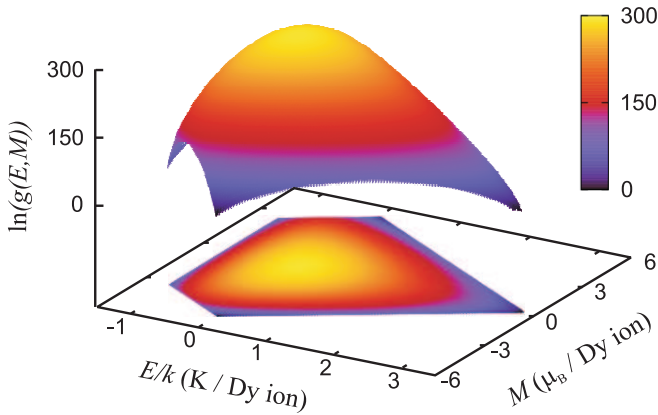}}
\caption{Logarithm of the DoS as a function of $E$ and $M$ along [111] calculated for a system of size $L=3$. The figure also shows the projection of the DoS over the $M - E$ plane. }
\label{DoSEM}
\end{figure}
%

Fig.\ \ref{DoSEM} shows the logarithm of the DoS as a function of $E$ and $M$ along [111] calculated for a system of size $L=3$.  As expected, the DoS is still asymmetric along the $E$ direction (cf.\ Fig. \ref{DoS}) and symmetric in the $M$ axis.  The figure also shows the projection of the DoS over the $M - E$ plane, which takes the shape of a pentagon.  In this projection one can see that while the highest energy state corresponds to $M=0$ the ground state manifold comprises a range of magnetisation values; for $H \parallel [111]$ this range goes from -3.33 to 3.33 $\mu_B$ which is the highest value of the magnetisation that can be obtained along [111] without breaking the ice rules.

\begin{figure}[ht]
\centerline{\includegraphics[width=\columnwidth]{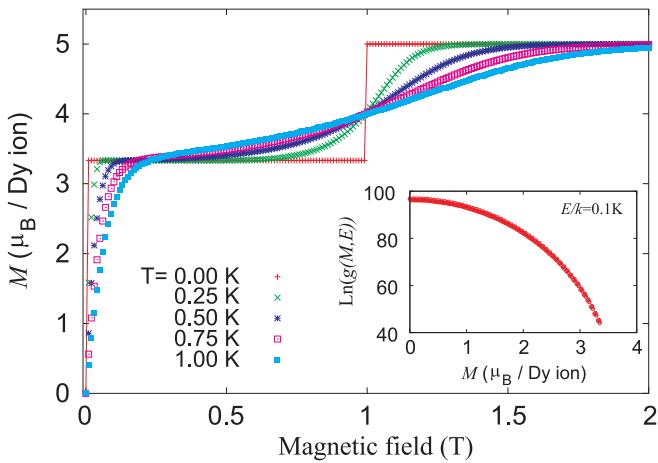}}
\caption{Magnetization vs field along the [111] axis for L = 3 and different values of temperature. The magnetization curves present two  well defined plateaux.  The first one, at $3.33 \mu_B /\mbox{Dy ion}$, correspond to a state where all apical spins are aligned with the field (the kagome-ice state). The second plateau, at $5 \mu_B / \mbox{Dy ion}$, corresponds to the state of maximal spin projection along [111].  The inset shows the logarithm of the density of states as a function of the magnetisation at a fixed energy ($k$ 0.1K above the ground state).}
\label{fig:MvsH111}
\end{figure}

In Refs \onlinecite{Moessner2003, Isakov2004} the case of the external field along [111] has been studied for the nnSI model both analytically and through conventional Monte Carlo simulations.  Along this crystallographic direction the system can be thought of as a stack of alternating kagome and triangular planes.  In the triangular planes the projection of the spins in the [111] direction is 1, while in the kagome planes it is 1/3.  The evolution of the system under field can be obtained by analysing the behaviour of the magnetisation at low temperatures.  In Fig.\  \ref{fig:MvsH111} we show $M$ as a function of the applied magnetic field as calculated using the WLA with two variables (energy and magnetisation) for $L=3$ and for the parameters of Dy$_2$Ti$_2$O$_7$, which is in perfect coincidence with all the features found in ref.\  \onlinecite{Isakov2004}.  At very low fields the magnetisation rises linearly with a slope which is proportional to $1/T$.  In this regime the field is merely selecting a subset with non-zero magnetisation along [111] from the ground-state manifold.  The slope is given by the competition between the gain in Zeeman energy and the entropy, because the number of states at a given magnetisation decreases sharply as $M$ increases.  This is  shown in the inset of Fig. \ref{fig:MvsH111} where the logarithm of the DoS is plotted as a function of the magnetisation at a fixed energy ($k$ 0.1K above the ground state). This shows how the logarithm of the number of available states is halved as the magnetisation is raised towards the $3.33 \mu_B /{\rm Dy~ion}$. When the field is increased to a value high enough to overcome the entropic effects, but low enough that the ice-rules are still obeyed, the spins in the triangular planes (with projection 1) all align with the field, the kagome planes decouple, and the system becomes effectively two-dimensional termed as ``kagome-ice'' (KI).  The spin-ice rules in the KI still allow for an exponentially degenerate number of configurations, and it still possesses an extensive residual entropy (as shown below).   The magnetisation reaches a plateau in this field range at $m=3.33 \mu_B$ per Dysprosium ion, which is easily calculated by taking into account the ice-rules and the different projections of the spins in the kagome and triangular planes.  If the field is increased further, it eventually overcomes the exchange interaction and the system goes through a metamagnetic transition (at around 1 Tesla in Fig.\ \ref{fig:MvsH111}) to a fully polarised state where all spins maximise their projection with the magnetic field ($m=5 \mu_B/{\rm Dy~ ion}$).  In the nnSI model this transition is merely a crossover and its width is strongly temperature dependent. If the dipolar interactions are added to the model, the transition becomes a first order \cite{Castelnovo2008}.

%
%

As mentioned before, the ice-rules in the KI phase are still under-constraining the system and allow for an exponentially large number of possible configurations.  In this case it is possible to obtain an exact solution for the residual entropy due to a mapping from the KI into dimers in a honeycomb lattice \cite{Moessner2003,Udagawa2002}.  This mapping allows the study of the effects of slightly misaligned fields, which lead to a two-dimensional Kasteleyn transition (see \onlinecite{Moessner2003}), as well as the transition to the fully polarised state, which can be interpreted as a dimer-monomer transition and is expected to be accompanied with a peak in the entropy as a function of field \cite{Isakov2004}.

In conventional MC simulations, the calculation of the field dependent entropy usually includes the integration of the specific heat as a function of temperature, with the integration constant for each field point determined by the value at an appropriate fixed point.  In the case of the WLA this calculation is straightforward from the known $g(E,M)$ and requires no further input.

\begin{figure}[ht]
\centerline{\includegraphics[width=\columnwidth]{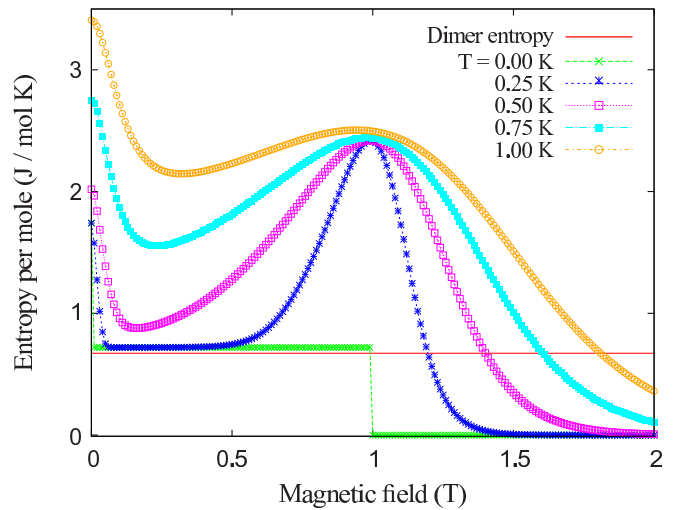}}
\caption{Entropy vs magnetic field for L = 3 and for different values of the temperature. The $T=0$ curve shows how the entropy is reduced to zero through two steps: a first one into the KI state, and a second one (at about 1 tesla) when the system becomes fully polarised.  At higher temperatures the most noticeable feature is a giant entropy peak at the polarisation transition (see text).}
\label{SvsH111}
\end{figure}

We have calculated the entropy as a function of field along [111] for different temperatures (see Fig.\  \ref{SvsH111}) from the DoS for $L=3$. The green curve shows the $T=0$ result: here the entropy jumps from its zero field value (the 3D ice-entropy) to the KI value (close to 0.7 J/mol K) and at higher fields (around 1 tesla) jumps down to zero as the system becomes fully polarised.  The red line is the exact value for the KI entropy ($S_0=0.6715$ J/mol K) as calculated in Refs.\   \onlinecite{Moessner2003,Udagawa2002}.  The slight discrepancy between the calculated $S_0$ and the exact result is due to finite size effects.  It is worth pointing out that despite its name and the fact that the coordination number of the lattice is 4, the KI is not {\em sensu stricto} an ice-type model and thus its residual entropy is smaller: $W_{KI}=1.175$, compared to $W_{2D}=(4/3)^{3/2}=1.539 \ldots$ . As the temperature increases, the most notable feature is the appearance of a giant peak in the entropy, which at low temperature (below 0.7K) is larger than the residual entropy.  It might seem counterintuitive that the application of an ordering field might result in an increase of the entropy.  This can be explained simply, with the aid of the dimer mapping, as coming from the crossing of an extensive number of energy levels, corresponding to different numbers of monomer defects, which have macroscopic entropies (see \onlinecite{Isakov2004}). In real materials though, this feature of the nnSI model, which has potential applications for magnetocaloric manipulations, is almost completely suppressed by additional interactions.

%
%

%
\begin{figure}[ht]
\centerline{\includegraphics[width=\columnwidth]{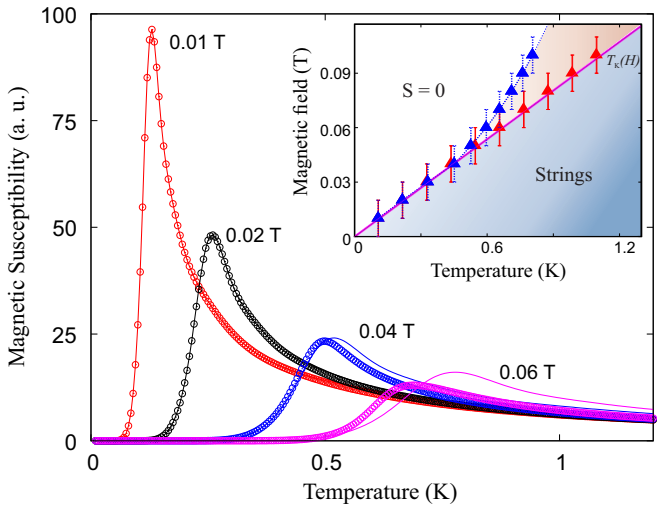}}
\caption{Linear magnetic suceptiblity $\chi$ as a function of temperature for fixed fields (circles).  The lines show the same magnetic susceptiblity calculated using solely configurations obeying the spin-ice rule (see text). The inset shows  the field dependence of the Kasteleyn transition $T_K$ as a function of temperature as extracted from $C_H$ and $\chi_H$.   The blue triangles correspond to those determined from the calculation of $C_H$ and $\chi_H$ using all states, while the red to those using spin-ice configurations only.  The solid line is the theoretical prediction of $T_K(H)$. }
  \label{ChivsT100}
\end{figure}
%


\subsection{Spin-ice: Field along [100] \label{h100}}

A case of particular interest arises when an external field is applied in the [100] direction.  In this case, contrary to the previous one, the fully saturated state belongs to the zero field ground-state manifold, that is, it satisfies the ice-rules.  At low temperatures ($kT \ll J_{\rm eff}$) and for any value of the magnetic field, there are no excitations in the form of local violations to the ice-rules.  In Ref. \onlinecite{Jaubert2008,Jaubert2009} it was shown that in this regime the competition between entropy gain and loss of Zeeman energy gives rise to a three-dimensional Kasteleyn transition \cite{Kasteleyn1963} where strings of negative magnetisation proliferate and span the whole length of the sample.  This transition takes place at a field dependent critical temperature given by \cite{Jaubert2008}
\begin{equation}
 T_K= \frac{ 2 \mu H_K }{\sqrt{3} k_B \ln 2},
 \label{TK}
\end{equation}
below which no string is present.  This critical temperature comes from equating in the free energy the loss in Zeeman energy per segment of a string of negative magnetisation ($2/\sqrt{3}h$, due to the spin projection) with the term arising from the entropic gain per segment ($T\ln 2$).  Since the line spans the whole sample, an equal number of in-pointing  and out-pointing spins are flipped and the ice-rule is preserved in the whole sample, that is to say, the Kasteleyn transition occurs between different spin-ice states.  In the case of our simulation (where we have imposed periodic boundary conditions) these lines take the shape of non-contractible closed loops in the torus.  Notice that the characteristic energy of this process at low temperatures is completely independent of  the value of $J_{\rm eff}$ and will be present even in the limit $J_{\rm eff}/kT \to \infty$.

The main charateristic of a Kasteleyn transition is its asymmetry: excitations are only possible at the disordered side of the transition.  This is shown in Fig.\ \ref{ChivsT100} where the magnetic susceptiblity is plotted as a function of temperature at a series of fixed fields as obtained from our WLA simulation for a system size of $L=3$ (circles).  There, it is clearly seen that at low temperatures, while the susceptiblity tends to diverge when $T_K $ is approached from the disordered side --resembling a second-order phase transiton-- it is flat on the other side --a behaviour more akin to a first-order phase transition.  This transition was initially  termed as ``3/2-order'' \cite{Nagle1973,Nagle1975} and later as ``K-type'' \cite{Nagle1989}.

%
\begin{figure}[ht]
\centerline{\includegraphics[width=\columnwidth]{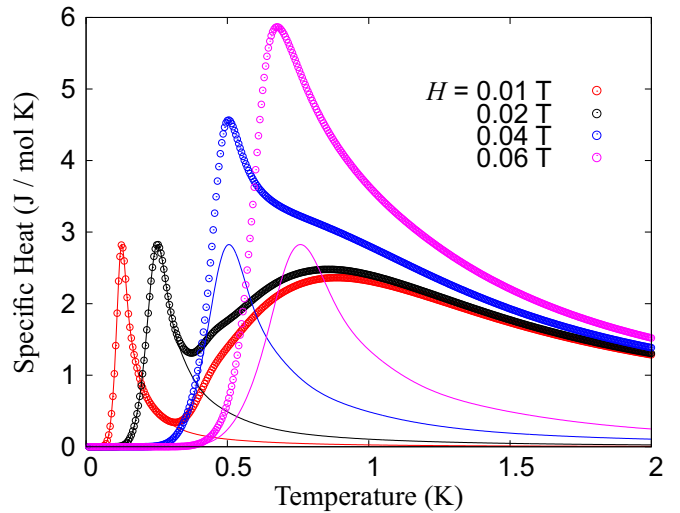}}
\caption{Specific Heat vs temperature for L = 4 and different values of field along [100] (circles). For low fields, it is easy to distinguish the high temperature Schottky-type peak characteristic of the onset of spin-ice correlations from the low temperature asymmetric peak due to the K-type transition.  As the field is raised, the transition moves to higher $T$ and it is gradually affected by the presence of additional local excitations. The lines show the specific heat calculated using only configurations obeying the ice-rule, consequently, the Schottky-type peak is absent.}
  \label{CvsT100}
\end{figure}
%

A similar asymmetric peak should be expected in the specific heat, $C_H$.  Fig. \ref{CvsT100} shows $C_H$ as a function of temperature for a series of fixed fields as extracted from our simulations.  At low fields, it is easy to distinguish two clearly defined peaks, the first one, at high temperatures corresponds to the Schottky-like peak already mentioned in the $H=0$ section corresponding to the onset of spin-ice correlations in the system.  The low temperature peak corresponds to the Kasteleyn transition, and shows the expected sharp edge at low temperatures and gradual rise on the high temperature side.  As the field is increased, the K-type transition is moved towards higher temperatures.  In our simulations $J_{\rm eff}/k = 1.11{\rm K}$  so the condition of $J_{\rm eff} \ll kT$ is no longer satisfied at $T_K(H)$ and point like excitations are seen at both sides of the transition, gradually changing the peak into a more symmetrical shape. As point excitations become more important, the simple argument sketched above for the field dependence of $T_K$ is no longer valid, and the transition widens and deviates from a linear dependence (see the blue triangles in the inset of Fig.\ \ref{ChivsT100}).

In the WLA, it is possible to impose additional constraints on the Hamiltonian by selectively choosing the states used to construct the partition function.  In this case in particular, it is very simple for finite size lattices to identify the states that strictly obey the ice-rule by their magnetisation value.  In this way, it is possible to calculate the different thermodynamic quantities for the ideal Kasteleyn transition, isolating the effect of point-like defects.  The results for the susceptiblity and specific heat are shown as solid lines in figures \ref{ChivsT100} and \ref{CvsT100} respectively.  They coincide at low temperatures, when $kT \ll J_{\rm eff}$ and the unconstrained curves show a transition at a lower temperature.  The most noticeable change is seen in the specific heat, where the Schottky-like peak is absent from the ice-rule obeying curves.  Furthermore, if we repeat the analysis to extract $T_K(H)$ from this latter set of data,  the curve (red triangles in the inset to Fig .\ \ref{ChivsT100}) follows very closely the theoretical prediction (solid line).

%
\begin{figure}[ht]
\centerline{\includegraphics[width=\columnwidth]{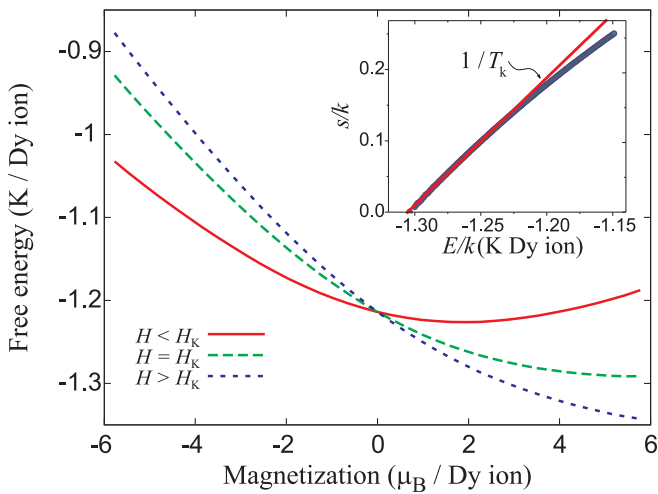}}
\caption{ The Gibbs potential for the system as a function of the magnetisation calculated using the WLA for ice-rule configurations at a fixed temperature $T=0.2K$ and for three different fields along [100]: $H_1  > H_K$ (blue symbols), $H_2=H_K$ (green symbols), and $H_3 < H_K$ (red symbols). The line is a guide to the eye.  The inset shows the entropy per spin, $s$ as a function of the energy for a fixed field of 0.05 T (blue dots).}
  \label{GvsM100}
\end{figure}
%

One of the unique characteristics of the WLA is that it makes it  simple to calculate the dependence of the free energy as a function of a chosen order parameter --in our case the magnetisation.  This provides valuable information regarding the nature of a phase transition, and is particulary interesting for an unusual case such as the K-type transition.

As we have mentioned before, the Kasteleyn transition takes place when $J_{\rm eff}/kT$ is small enough that excitations that break the ice-rule are extremely improbable.  In this case, the  energy of the system is a constant, the free energy to a purely entropic term and the Gibbs potential is given by $G=-TS-MH$.  This resembles the case of a simple paramagnet, however, as discussed by Jaubert and collaborators (see \onlinecite{Jaubert2009Proc,JaubertPhD2009}) a crucial difference arises from the ice-rule constraint:  if, contrary to the case of the paramagnet, this constraint brings the entropy to zero at a {\em finite} $H/kT$, this is sufficient to drive a Kasteleyn transition in the system.  This ad-hoc supposition can be put to test using the WLA.  The inset of Fig.\ \ref{GvsM100} shows the behaviour of the entropy per spin, $s$ as a function of the energy in the neighbourhood of $s=0$ for a fixed field of 0.05 T.  As seen in the figure, the slope at which the entropy vanishes is indeed finite and, furthermore, it is given precisely by $1/T_K$, with $T_K$ the Kasteleyn temperature for this field determined from $\chi$ and $C$.

The main panel of the figure shows the Gibbs potential as a function of the magnetisation, $G(M)$, at $T=0.2K$ for three different fields along [100]: $H_1 > H_K$, $H_2=H_K$ and $H_3 <H_K$ as determined using the WLA for a system of $L=4$ using only ice-rule configurations.  This figure captures the characteristic features expected for a K-type transition.  The low field curve resembles that of a paramagnet, with a wide minimum at a non-zero magnetisation.  As the field is raised towards $H_K$ this minimum becomes wider and moves to higher values of $M$,while fluctuations increase. At the critical field $H_K$, the system becomes singular, the minimum sits at  $M_{\rm sat}$ and the curve becomes flat ($dG/dM=0$) in its neighbourhood.  For $H>H_K$, the absolute minimum sits at $M_{\rm sat}$, the neighbourhood to the minimum is linear, with $dG/dM$ finite and negative, showing the complete absence of fluctuations in the ordered state.

\section{Conclusions}

In this work we have explored by means of the WLA the nearest-neighbour spin-ice model,
an example of a simple classical frustrated model.  We have determined the value of the residual entropy $S_0$ by doing a finite size analysis of $S_0(L)$ which can be calculated directly from the DoS determined by the WLA for samples of size $L^3$.  By including the magnetisation as one of the parameters of the DoS, we were able to calculate the thermodynamic properties of the nnSI model as a function of field.  In particular, we have investigated the cases where the field was applied in the [111] and the [100] directions.  In the first case, we demonstrated that the magnetisation as a function of field calculated with the WLA had all the features characteristic to this direction, namely, a $1/T$ rise to the kagome-ice plateau at 3.33 $\mu_B/{\rm Dy~ion}$ and a sudden jump at $H \approx 1 tesla$ to the fully saturated state.  The method allowed for a direct calculation of the field dependent entropy, $S(H)$, without the need of any additional fixed parameter.  As expected, $S(H)$ has a plateau at the KI phase with a value that tends to that determined by the mapping of the system into dimer configurations on a honeycomb, and shows a marked peak at the polarisation transition.  In a similar fashion, we showed that when a field is applied along [100],  a Kasteleyn transition takes place between a $S=0$ state and one where line-like excitations pierce the system.   Additionally, the WLA provides information not obtained through other methods, such as the free energy as a function of the order parameter, which, to our knowledge, has been calculated for the first time for a Kasteleyn transition. By this mean, we were also able to prove that the  ad-hoc assumption that the local constraint of the spin-ice rule, brings the entropy to zero at a {\em finite} $H/kT$, is  perfectly valid.  We have, furthermore, showed that the WLA allows the computation of the thermodynamic properties of the system when additional constraints --not present in the Hamiltonian-- are put into the system.  In particular, we showed that by selecting only the ice-rule states, we could calculate the behaviour of the ideal Kasteleyn transition, that is, the one that takes place in the absence of point defects.  In summary, we show that the WLA is a very useful tool for simulations of frustrated magnetic systems and provides accurate information for the thermodynamic properties in equilibrium. As a result, quantities such as the entropy and the free energy, that is cumbersome to obtain through other methods, can be easily computed. In addition, the algorithm can be used to study the system under the influence of additional constraints.

\begin{acknowledgments}
We thank R. Moessner and T.S. Grigera for helpful discussions. This
work was supported by Consejo Nacional de Investigaciones Cient\'{\i}ficas y T\'ecnicas (CONICET), Agencia Nacional de Promoci\'on Cient\'{\i}fica y Tecnol\'ogica (ANPCyT), Argentina and the Helmholtz Virtual Institute ``New states of matter and their excitations'', Germany.

\end{acknowledgments}

\bibliography{science2}

\end{document}